# Heatons induced by attosecond laser pulses

1 Janina Marciak-Kozlowska
2 Miroslaw Kozlowski

(1 Institute of Electron Technology
Al. Lotnikow 32/46, 02-668 Warsaw, Poland
2 Physics Department, Warsaw University
Hoza 69, Warsaw)

**Abstract**

In this paper the dynamics of the interaction of attosecond laser pulses with matter is investigated. It will be shown that the master equation: modified Klein-Gordon equation describes the propagation of the heatons. Heatons are the thermal wave packets. When the duration of the laser pulses is of the order of attosecond the heaton thermal wave packets are nondispersive objects. For infinite time the heatons are damped with damping factor of the order of relaxation time for thermal processes.

1. Introduction

The breakthrough progress has been made recently in the generation and detection of attosecond laser pulses with high harmonic generation techniques. This has heralded an age of attophysics in which many electron dynamics will be probed in real time.

In this paper the theoretical framework for attosecond laser pulses inter-action with matter is developed. In the set of papers in *Lasers in Engineering* and in the monograph: *Thermal processes using attosecond laser pulseses* we formulated the quantum hyperbolic thermal equation and find out its solution in different, important from technological point of view areas.

The basic notion of the quantum of the thermal field, the heaton was introduced in our papers. Up to now the *heatons* wave described in the static manner, without the investigation the dynamics of the heaton moving. In this paper based on the modified Klein-Gordon equation (MK-G) we investigate the motion of free heatons. As the result of the study of the solution of MK-G it will be shown that the *heatons* are the thermal wave packets. For the electron gas the *heatons* represents the packets with maximum coincident with the position of the electron in the space.

2. Physics at the attosecond frontier

The road from picosecond in femtosecond light pulses has been seen laser technology evolve towards lasers that emit light with greater spectral width – that is covering a wide of wavelengths. A short pulse results when all the spectral components in light beam interfere in a way that adds up to a single burst of light. The duration of this pulse is inversely proportional to the spectral width. This approach reach its natural limit when the spectral span of laser become a significant portion of the visible spectrum. This is because the pulse cannot be shorter than a period.

The latest advances are based on research into high-order harmonic of femtosecond laser pulses. Precise analysis of the way the gas atoms interact with the laser field requires careful application of quantum mechanics. The model of P. Corkum [1], the original laser pulse tears an electron away from an atom and the freed electron moves in response to the laser field being accelerated and decelerated as the electromagnetic field oscillates. The new harmonics are generated when the electron collides with the ion is left behind. This radiation is termed harmonic because its frequencies are multiples of the original laser frequency. P. Corkum argued that electrons that are ejected precisely at the peaks or the crests of the optical pulse are much like to radiate. So the radiation is produced in very short burst, which occupy just a fraction of the optical cycle. The bursts are estimated to last some 100 as. In paper [2] M. Hentschel et al., start with a 7-femtosecond optical pulse and estimate that after filtering more than 90% of the energy of the new radiation they produce is contained in a single 650 attosecond pulse.

The work of Hentschel et al. [2] hints at the likely direction that attophysics i.e. physics with attosecond laser pulses will take in the near future. We have now entered the attosecond world, but we will need a better guide-book to help us find the road of development.

In monograph [3] the theoretical framework for attosecond laser pulse interaction with matter was formulated. It was shown that when the laser pulse with duration of the order of attosecond interacts with matter it creates the transport processes which are described by hyperbolic transport equations.

When the duration of the laser pulse is shorter than the relaxation time $\tau$, $\tau = \frac{\hbar}{m\alpha^2 c^2}$ ($m$ = electron mass, $c$ is the light velocity and $\alpha$ is the fine structure constant, than the transport of thermal energy is described by the equation [4]

$$\frac{1}{v^2}\frac{\partial^2 T}{\partial t^2} + \frac{m_e \gamma}{\hbar}\frac{\partial T}{\partial t} + \frac{2V m_e \gamma}{\hbar^2}T - \frac{\partial^2 T}{\partial x^2} = F(x,t). \qquad (1)$$

In equation (1) $T$ is temperature, $m_e$ is electron mass, $V$ is the potential, $\gamma = (1-\alpha^2)^{-\frac{1}{2}}$ and $F(x,t)$ is the external force.

The solution of equation (1) can be written as

$$T(x,t) = e^{-\frac{t}{2\tau}} u(x,t), \tag{2}$$

where $\tau$ is the relaxation time, $\tau \cong 100$ as i.e. is the order of time duration of attosecond laser pulse. After substituting formula (2) to Eq. (1) we obtain

$$\frac{1}{v^2}\frac{\partial^2 u}{\partial t^2} - \frac{\partial^2 u}{\partial x^2} + qu(x,t) = e^{\frac{t}{2\tau}} F(x,t). \tag{3}$$

and

$$q = \frac{2Vm}{\hbar^2} - \left(\frac{mv}{2\hbar}\right)^2$$

$$m = m_e \gamma.$$

In the subsequent we will consider the relativistic electrons, without the external forces, i.e. for $F(x,t) \to 0$. In that case Eq. (3) can be written as

$$\frac{1}{v^2}\frac{\partial^2 u}{\partial t^2} - \frac{\partial^2 u}{\partial x^2} - \left(\frac{mv}{2\hbar}\right)^2 u(x,t) = 0. \tag{4}$$

3. Thermal wave packets induced by attosecond laser pulses

Equation (4) is the modified Klein-Gordon equation which can be written as

$$\Box u - \left(\frac{mv}{2\hbar}\right)^2 u(x,t) = 0, \tag{5}$$

where d'Alembert operator $\Box$ is equal

$$\Box = \frac{1}{v^2}\frac{\partial^2}{\partial t^2} - \frac{\partial^2}{\partial x^2}. \tag{6}$$

The ordinary Klein-Gordon equation for the particle with mass $m$ is of the form [5]

$$\Box = u + \left(\frac{m_0}{2\hbar}\right)^2 u = 0. \tag{7}$$

Equation (5) can be split into its real and imaginary parts. Putting for $u(x,t)$

$$u(x,t) = \Re(t,x)\exp\left[\frac{i}{\hbar}S(t,x)\right]$$

one obtains

$$\eta^{ab}(\partial_a S)(\partial_b S) = \hbar^2 \frac{\Box\Re}{\Re} - \left(\frac{mv}{2}\right)^2, \qquad (8)$$

where

$$\eta_{ab} = \text{diag}(1, -1, -1, -1), \qquad a, b = 1, 2, 3, 4.$$

We use Mackinnon's suggestion [6] therefore we look for solutions that satisfy the equation

$$\frac{\Box\Re}{\Re} = \left(\frac{mv}{\hbar\sqrt{2}}\right)^2. \qquad (9)$$

In this way Eq. (8) becomes

$$\eta^{ab}(\partial_a S)(\partial_b S) = m^2 v^2 = P_\mu P^\mu. \qquad (10)$$

If the velocity $v$ is constant, we have from (8)

$$S = -P^\mu P_\mu \qquad (11)$$

and the de Broglie relation $P^v = \hbar K^v$ holds where $K^v$ is the wave number and $P^v$ is the classical relativistic four momentum, $P^v = \left(\frac{E}{v}, \vec{p}\right)$. In paper [6] L. Mackinnon constructs a wave packet considering that a wave $\Phi = \Phi_o \exp[i\omega t]$ of frequency $\omega = \frac{mc^2}{\hbar}$ is associated with a particle of rest mass $m$ and that for an observer moving with a constant velocity $v$ with respect to the particle, the associated wave (by means of a Lorentz transformation) acquires the form $\Phi = \Phi_o \exp\left[i\omega\gamma\left(t - \frac{v}{c^2}x\right)\right]$. Mackinnon showed the wave packet is compatible with the basic experiments of quantum mechanics and does not spread in time the found

$$\Re = A\frac{\sin[gr]}{gr}, \qquad (12)$$

where $A$ = constant and $g = \frac{mv}{\hbar\sqrt{2}}$, and $g = \frac{mac}{\hbar\sqrt{2}}$ and

$$r = \gamma(x - vt) \tag{13}$$

is the distance from the particle portion, so that

$$u(x,t) = A \frac{\sin[gr]}{gr} \exp\left[-\frac{i}{\hbar}(Et - px)\right] \tag{14}$$

is the Lorentz boost of the solution

$$\Phi = A \frac{\sin gr'}{gr'} \exp[-i\omega t'],$$

where $r' = x$. This solution was first found by de Broglie [7] and also used in the stochastic interpretation of quantum mechanics by Vigier and Gueret [8].

In the monograph [3] the program of the quantization of the temperature field was undertaken. The master quantum equation for the temperature field was equation (5). The quantum of temperature field, the *heaton* was introduced. The *heaton* energy $E = \hbar\omega$, with $\omega = \frac{mv^2}{\hbar}, \omega = \frac{m\alpha^2 c^2}{\hbar}$. Considering equation (5) the *heaton* is the thermal wave packet which maximum coincides with electron and which shape is described by equation (14).

## 4. Conclusions

In this paper the modified Klein-Gordon (MK-G) equation for temperature field was obtained. It was shown that the solution of MK-G equation is the *quantum* of the temperature field, *heaton* (introduced in our paper *Found. of Physics Letters,* **9**, (1996) p. 235). The *heaton* is the thermal wave packet which maximum coincides with electron and which is moving with velocity $v = \alpha c$. As the result of the interaction of the attosecond laser pulses with electron gas the *heatons* are created. When the laser pulse is shorter than the relaxation time $\tau = \frac{\hbar}{m\alpha^2 c^2}$ the *heatons* are not dispersed.

References

[1]     P. B. Corkum, *Phys. Rev. Lett., 71,* (1993), p. 1994.

[2]     M. Hentschel et al., *Nature,* 414, (2001), p. 509.

[3]     M. Kozlowski, J. Marciak-Kozlowska, *Thermal processes using attosecond laser pulses,* Springer, 2006

[4]     J. Marciak-Kozlowska, M. Kozlowski, *Lasers in Engineering,* 12, (2002), p. 17.

[5]     W. A. Strauss, *Partial* Differential *Equation. An Introduction,* J. Wiley, N.Y. 1992.

[6]     L. Mackinnon, *Letter at Nuovo* Cimento, 32, (1981), p. 311.

[7]     L. de Broglie, C. *R. Acad. Sci.,* 180, (1925), p. 498.

[8]     J. P. Vigier and Ph. Gueret, *Found. Phys.,* 12, (1982), p. 1057.